\documentclass[sigconf,nonacm]{acmart}

\usepackage{gensymb}
\usepackage{array}
\usepackage{dblfloatfix}    
\usepackage[absolute,overlay]{textpos}

\AtBeginDocument{%
  \providecommand\BibTeX{{%
    \normalfont B\kern-0.5em{\scshape i\kern-0.25em b}\kern-0.8em\TeX}}}

\begin{document}

\title[Vacuum-formed 3D Printed Electronics]{Vacuum-formed 3D printed electronics: fabrication of thin, rigid and free-form interactive surfaces}

\author{Freddie Hong}
\email{t.hong19@imperial.ac.uk}
\orcid{https://orcid.org/0000-0002-6318-9723}
\affiliation{
  \institution{Imperial College London}
  \streetaddress{Imperial College Road}
  \city{London}
  \state{London}
  \postcode{SW7 1AL}
}

\author{Luca Tendera}
\email{ltendera@student.ethz.ch}
\orcid{}
\affiliation{
  \institution{ETH Zurich}
  \streetaddress{Raemistrasse 101}
  \city{Zurich}
  \state{Switzerland}
  \postcode{8092}
}

\author{Connor Myant}
\email{connor.myant@imperial.ac.uk}
\orcid{https://orcid.org/0000-0003-4705-4009}
\affiliation{
  \institution{Imperial College London}
  \streetaddress{Imperial College Road}
  \city{London}
  \state{London}
  \postcode{SW7 1AL}
}

\author{David Boyle}
\email{david.boyle@imperial.ac.uk}
\orcid{}
\affiliation{
  \institution{Imperial College London}
  \streetaddress{Imperial College Road}
  \city{London}
  \state{London}
  \postcode{SW7 1AL}
}

\begin{abstract}

Vacuum-forming is a common manufacturing technique for constructing thin plastic shell products by pressing heated plastic sheets onto a mold using atmospheric pressure. Vacuum-forming is ubiquitous in packaging and casing products in industry spanning fast moving consumer goods to connected devices. Integrating advanced functionality, which may include sensing, computation and communication, within thin structures is desirable for various next-generation interactive devices. Hybrid additive manufacturing techniques like thermoforming are becoming popular for prototyping freeform electronics given its design flexibility, speed and cost-effectiveness. In this paper, we present a new hybrid method for constructing thin, rigid and free-form interconnected surfaces via fused deposition modelling (FDM) 3D printing and vacuum-forming. While 3D printing a mold for vacuum-forming has been explored by many, utilising 3D printing to construct sheet materials has remains unexplored. 3D printing the sheet material allows embedding conductive traces within thin layers of the substrate, which can be vacuum-formed but remain conductive and insulated. We characterise the behaviour of the vacuum-formed 3D printed sheet, analyse the electrical performance of 3D printed traces after vacuum-forming, and showcase a range of examples constructed using the technique. We demonstrate a new design interface specifically for designing conformal interconnects, which allows designers to draw conductive patterns in 3D and export pre-distorted sheet models ready to be 3D printed.
\end{abstract}

\keywords{3D printed electronics, conductive filament, hybrid additive manufacturing}

 \begin{teaserfigure}
   \includegraphics[width=\textwidth]{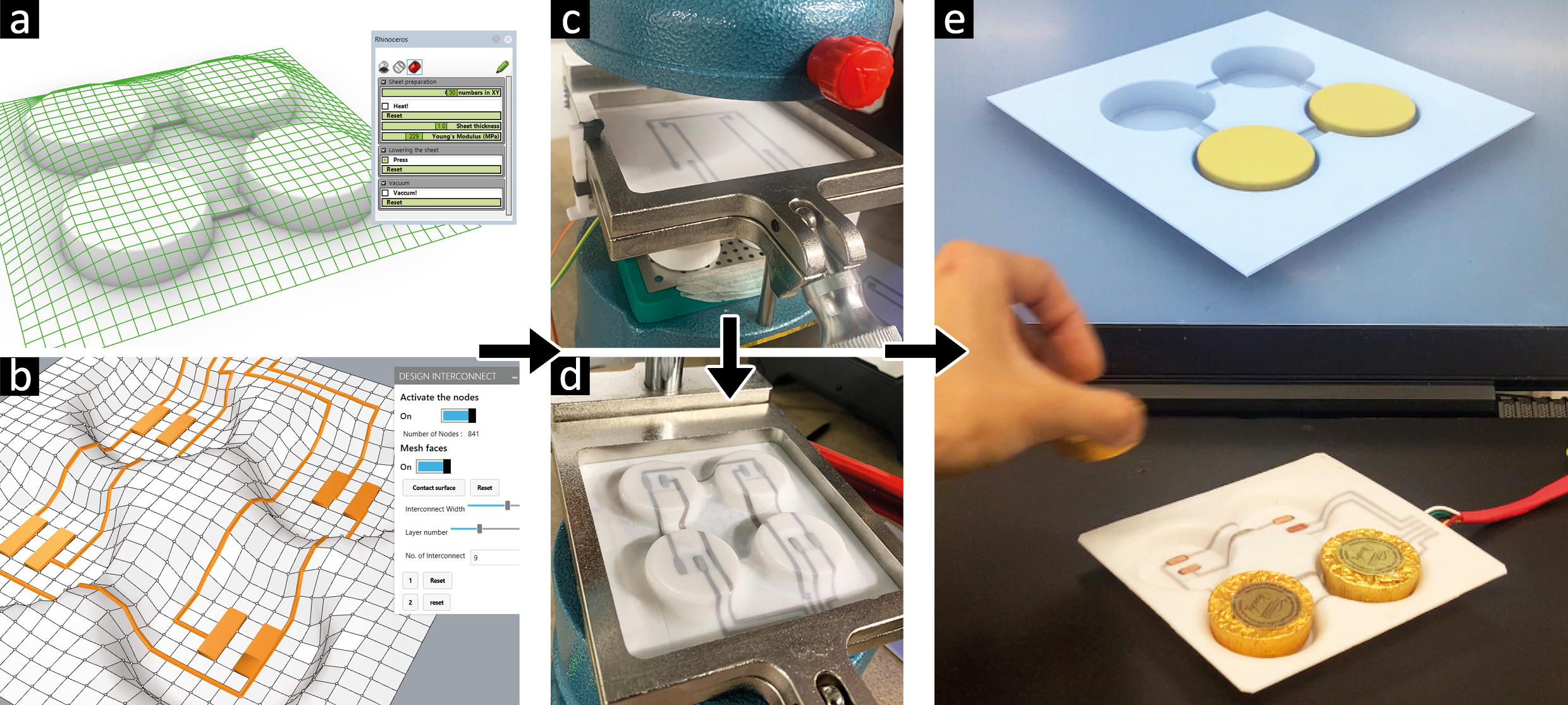}
   \caption{Overview of a vacuum-formed smart surface. a) Design editor to simulate vacuum-forming; b) Design tool for drawing interconnects on a vacuum-formed surface; c) Pre-distorted 3D printed sheet clamped to low-cost vacuum-forming machine; d) Vacuum-formed 3D printed sheet; e) assembled \& connected smart tray sensing the presence of an object}
  \Description{
  Overview of the fabrication process (a) is a CAD software screen showing the simulation process including the control parameters and 3D geometry. (b) is a post-simulation CAD image with conductive interconnected being design onto the surface. (c) is a vacuum-forming machine with the 3D printed sheet clamped onto. (d) shows a 3D printed sheet vacuum-formed into a tray shape. (e) shows a working smart tray sample connected to the computer which shows how many chocolate is placed in the tray.}
   \label{fig:teaser}
 \end{teaserfigure}

\maketitle 

\section{Introduction}

Thin, rigid 3D surfaces with integrating conductive interconnects may be a desirable structures for the next generation of smart devices. Interconnected 3D devices open up a range of new design opportunities for displays, wearables and interactions. The two most popular approaches for manufacturing such rigid interconnected devices are: i) molded interconnect device (MID)~\cite{Islam2009ProcessDevices}, which involve injection molding the 3D object, laser structuring the surface and electroless plating the activated traces with copper, nickel and gold, and ii) in-mold electronics (IME), which involve 2D printing conductive silver patterns on polymer-film that is thermoformed and transferred onto the surface of the mold \cite{NextFunctionalCenturies}. While these methods create highly accurate and functional electronic device, they are best suited to mass-production owing to the high cost of tooling, molding requirement and limited design flexibility inherent to the multi-step manufacturing processes.

As a result, more flexible personal fabrication approaches to constructing conformal circuits have been explored such as hydroprinting conductive pattern onto 3D objects \cite{Groeger2018ObjectSkin:Touch,Ng2019ConformalSurfaces,Saada2017HydroprintingStructures}, adhering inkjet printed circuits~\cite{10.1145/3379337.3415898,10.1145/2556288.2557150} and copper tapes to object surfaces \cite{Savage2012Midas:Objects,Umetani2017SurfCuit:Prints}. These methods use inexpensive commercial equipment like ink-jet printers, and are easily accessible for a wide range of potential users. Since the 3D printed substrate and the 2D inkjet printed circuit are prepared separately, the aforementioned methods are less capable of constructing double-sided or embedded circuits, and require additional manual labor. 

Researchers have also 3D printed conductive polylactic acid (PLA) to construct various sensing elements~\cite{Schmitz2015Capricate:Objects,10.1145/3290605.3300684,10.1007/978-3-319-22701-6_25}. However, due to the planar construction of layer by layer deposition, the 3D printed interconnects have inconsistent print quality in the vertical direction, in addition to poor conductivity that limits the application space of conductive PLA. While the recent method of copper electroplating the conductive PLA by Angel et al. \cite{Angel2018} and Kim et al. \cite{Kim2019} can reduce the resistance of the printed conductive PLA trace significantly, poor print quality of the base trace can cause unevenness in electroplating. 

In this paper, we present a novel hybrid additive manufacturing technique of vacuum-forming 3D printed sheets to construct various conformal electronics. We first 3D print various sheet materials consisting of PLA and conductive PLA using a low-cost desktop 3D printer, then we employ a desktop vacuumforming machine to vacuum-form 3D printed sheet material into a desired shape. While 3D printing molds have been explored by many\cite{Hartman2014BenefitsMolds,FormlabPrototypingMolds}, the approach of using 3D printed sheets with vacuum-forming remains largely unexplored.

Key benefits of 3D printing the sheet materials include: i) a user can vary the thickness of the sheet within a single material and mix multiple materials together in a single sheet (e.g. colors and functions), ii) it requires significantly less time to construct free-form geometries compared to traditional 3D printing, and iii) the vacuum formed structures have a continuous layer structure that has beneficial mechanical properties and can enhance the conductivity of the printed trace.
We also share our design editor for designing the vacuum-formed electronics which includes vacuum-forming simulation integrated within a common 3D CAD environment, graphical interface for drawing the interconnects and pre-distorting the final design into a sheet model ready to be exported for 3D printing. Our contributions in this paper are summarized as follows:

\begin{itemize}
    \item We demonstrate a novel fabrication technique for 3D printed sheets to construct thin, freeform and rigid interconnected surface using only inexpensive and accessible equipment and materials exploiting vaccum-forming, a simplified type of thermoforming~\cite{Hong2020ThermoformedBending}
    \item We analyse vacuum-foming behaviour of 3D printed sheet and evaluate electrical properties of vacuum-formed 3D printed traces
    \item We provide a parametric design editor\footnote{Our software and additional `how-to' information will be made openly available on GitHub upon this paper's acceptance.} integrated within a 3D modelling environment to aid the user to model the vacuum-formed interconnected surface with automated 3D to 2D mapping.
    \item We demonstrate the capabilities of vacuum formed 3D printed surface by prototyping example applications with various form factors and functional characteristics.
\end{itemize}

\section{Related work}

Our research spans the fields of hybrid additive manufacturing and personal fabrication of electronic devices in HCI.

\subsection{Fabricating conformal electronic devices}
Various approaches for integrating conductive traces into 3D objects have been introduced, comprising mainly two methods; i) translating separately prepared flat conductive patterns onto 3D surfaces, and ii) constructing the conductive pattern directly onto the 3D surface. In the first case, Saada et al. \cite{Saada2017HydroprintingStructures} and ObjectSkin \cite{Groeger2018ObjectSkin:Touch} use a hydroprinting technique to translate 2D inkjet-printed silver ink patterns onto a 3D surface via water-soluable polyvinyl alcohol (PVA) film. In a similar manner, Ng et al. \cite{Ng2019ConformalSurfaces} screen print graphene patterns onto PVA film and translate them onto 3D surfaces also using the hydroprinting method. Recently, MorphSensor \cite{Zhu2020MorphSensorModules} showcased a method of inkjet printing silver patterns onto adhesive tape to integrate circuits onto 3D surfaces. While these methods successfully construct conformal circuit onto arbitary surfaces, they require a separate construction of the substrate and the conductive patterns, which are inefficient for constructing double-sided circuits in addition to the added manufacturing steps involving manual labor. Recently, Hong et al. introduced Thermoformed Circuit Boards (TCB) \cite{Hong2020ThermoformedBending}, a technique of 3D printing both conductive PLA and substrate PLA in a single manufacturing step that is followed by heat bending. As TCB constructs both elements in a single print, it can conveniently fabricate double-sided circuits. As is inherent with folding flat material, however, TCB is limited in form factor, and in terms of accuracy and repeatability because of fabrication elements `by hand'.

To construct the circuit directly on the 3D surface, Plain2Fun \cite{Wang2018Plain2Fun:Circuits}, for instance, mounts a conductive ink pen on a 4 degree of freedom stage to draw conductive patterns on an object, Aerosol Jet system by Optomec \cite{OptomecElectronics} uses droplet-based deposition of electronic ink on multi-axis system to print complex conformal circuits on 3D substrates, and similarly, Adams et al. \cite{Adams2011} use extrusion of conductive silver ink on a 3-axis positioning stage to construct electrically small antennas on hemispheric substrates. A more manual method, SurfCuit~\cite{Umetani2017SurfCuit:Prints}, has shown the method of 3D printing cavities on freeform objects and mounting conductive copper tape to form the circuit on surface, and Midas \cite{Savage2012Midas:Objects} also showed a similar method of attaching copper foil conformally onto a 3D object for fabricating various touch sensors.
These methods of direct printing conductive patterns onto a substrate require high-precision multi-axis stages, which require complex toolpath preparation. Multi-axis equipment is also inaccessible for many individual designers.

In this paper, we propose a method of translating 3D printed sheet material to construct thin freeform, double-sided and interconnected devices. We showcase an approach of vacuum-forming prefabricated flat 3D printed sheets with conductive PLA traces to construct freeform devices that have higher degree of accuracy and repeatability. 

\subsection{Thermoforming 3D objects for personal fabrication}

Several methods of thermoforming sheet materials to prototype 3D objects have been introduced in the HCI community. Since thermoforming can construct 3D geometries quickly with minimal manufacturing steps, researchers have come up with various tool kits to explore this benefit in personal fabrication. For example, ProtoMold \cite{10.1145/3025453.3025498} and DrawForming \cite{10.1145/2839462.2856534} use shape changing mold beds consisting of actuators to eliminate the mold-making process and turn vacuum-forming in to a more intuitive craft. LaserOrigami~\cite{Mueller2013LaserOrigami:Objects} uses laser cutters not only to cut the shape, but also to thermoform acrylic sheets by heat-folding and stretching, thus minimising the manufacturing process by eliminating the assembly. Researchers have also developed user-interfaces for computationally thermoforming color textures onto 3D objects. Computational Thermoforming~\cite{Schuller2016ComputationalThermoforming} and Zhang et al.~\cite{Zhang2017ColoringThermoforming} showcased an interface that converts a textured surface of a 3D digital model into a pre-distorted 2D image ready to be printed on the sheet prior to vacuum-forming.

Our approach is similarly motivated to reduce manufacturing processes and accelerate the prototyping process by vacuum forming 3D printed 2D sheets in 3D freeform geometry. We also introduce our new user interface integrated within popular 3D CAD software to help users to design 3D interconnects in a parametric manner with an automated 3D to 2D distortion.

\subsection{Thermoforming stretchable circuits}

Preparing both the substrate and the conductive pattern in planar sheets prior to forming them into 3D objects allows designers to embed the conductive patterns within the substrate, bypass the complex tooling required in conformal consturction, and efficiently construct conformal interconnects for arbitrary concave and convex geometries. This method of thermoforming electronics relies on the `stretchability' of the traces. A common method of constructing stretchable circuits is by adopting wavy patterns or using stretchable conductive material. For instance, ElectroDermis \cite{10.1145/3290605.3300862}, Yang et al. \cite{Yang20183DTechnology}, and Huang et al. \cite{Huang2018Three-dimensionalElectronics} use the meandering pattern to comply with distortions and stretches without breaking the trace. Alternatively, Nagels et al. \cite{10.1145/3173574.3173762} use stretchable liquid metal embedded inside the silicone substrate to construct various stretchable devices. Devaraj et al.\cite{Devaraj2019} and IMold \cite{Ting2020IMold:Electronics} use conductive silver ink which remains connected after thermoforming. TCB~\cite{Hong2020ThermoformedBending} 3D prints conductive PLA, which has low glass transition temperature and low tensile modulus that can be distorted and stretched into various shape and remain connected.
In this work we also employ conductive PLA to construct 3D printed sheets with embedded conductive traces. 3D printing conductive PLA is based on low-cost and highly accessible equipment that is ideal for individual designers to employ in their home environment.

\section{Vacuum-forming 3D printed sheets}

While 3D printing the mold for vacuum-forming has been explored by many researchers \cite{Hartman2014BenefitsMolds}, a method using a 3D printer to fabricate a sheet material for vacuum-forming is largely unexplored. Eksi et al.'s analysis on deformation behaviour of vacuum-formed 3D printed PLA sheets was the first and only to characterise the thermoforming behaviour of the PLA sheet~\cite{Eksi2020ThermoformingSheets}. Here, we utilise a low-cost desktop vacuum-forming machine (JT-18, JINTAI) which has a maximum sheet dimension of $130\times130$ mm, and an open source FDM 3D printer (Prusa i3) to construct various conformal electronics. We used conventional white PLA (GEETECH), various colored PLA and conductive PLA (Multi3D), all of which are commercially available online internationally.

\subsection{Temperature and sheet thickness}

\begin{figure}[h]
  \centering
  \includegraphics[width=1\linewidth]{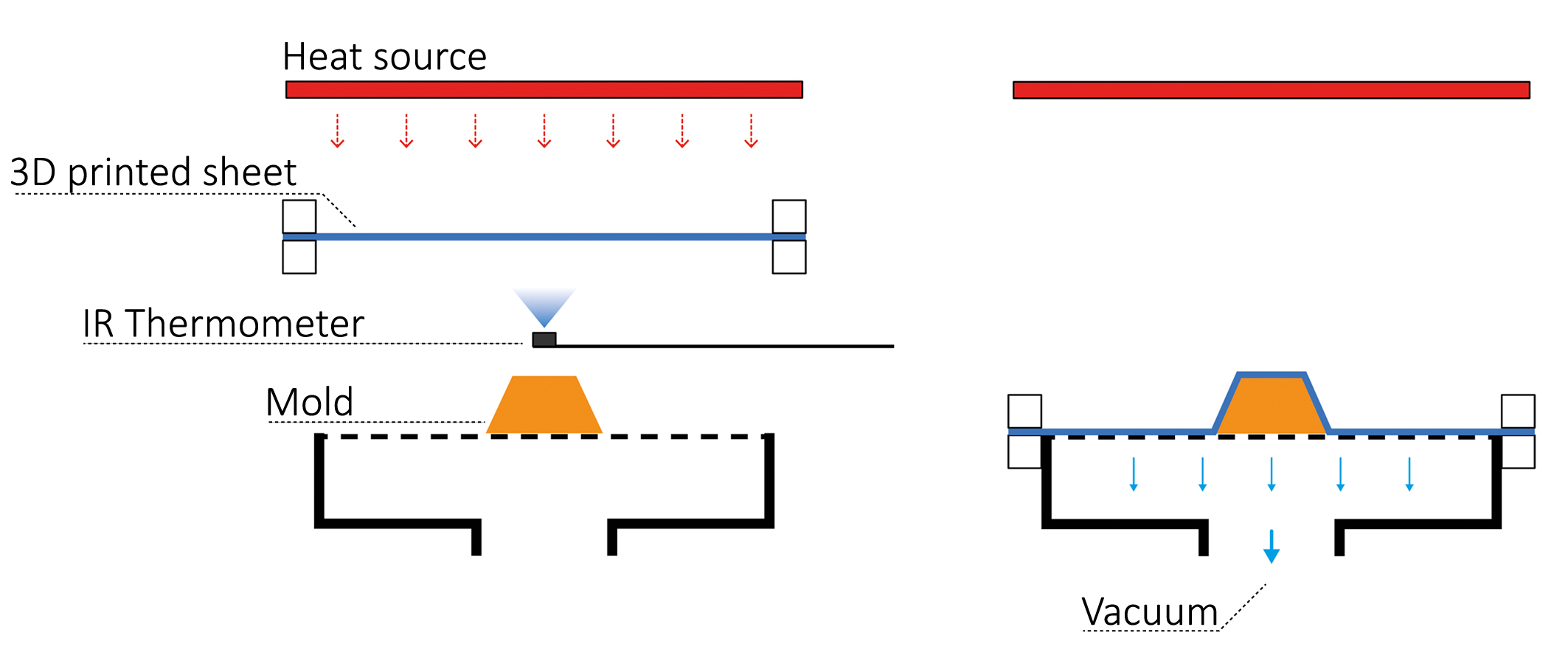}
  \caption{Schematic diagram of the vacuum-forming process. Left) 3D printed sheet is heated up until it reaches the forming temperature measured using IR thermometer; Right) 3D printed PLA sheet is lowered to the mold and vacuumed}
  \label{fig:vacuumform diagram}
  \Description{
  This figure shows an illustration of the vacuum-forming machine in cross-section. (Left) shows a 3D printed sheet clamped on to the machine with the infra-red thermometer positioned right below. It also shows a mold right below the thermometer and on top of the vacuum-forming bed. (right) shows a vacuum-formed sheet that is now positioned onto the mold. Both illustration shows a heater at the top.
  }
\end{figure}

To evaluate the vacuum-forming characteristics of the 3D printed sheets, we 3D printed 9 plain square sheets sized 130~mm by 130~mm with varying thicknesses of 0.5~mm, 1.0~mm and 1.5~mm. We then vacuum formed the sheets at varying forming temperatures of 110 \degree C, 120 \degree C, and 130 \degree C. The temperature was measured using an infra-red thermometer mounted underneath the sheet (Figure \ref{fig:vacuumform diagram}). Following the general guideline for vacuum-forming, we heated the 3D printed sheet until a visible sag was observed, indicating that the sheet is ready to be vacuum formed. In our experiments we found the the 3D printed PLA sheet began to sag at 110 \degree C. While the forming temperature may differ depending on the filament manufacturer and tensile properties, we found that most commercial PLA started to sag at 110 \degree C, owing to the low glass transitional temperature of the PLA (T\textsubscript{g}~< 60{\degree}C). The Young's modulus of the PLA, which measures the stiffness of the material, also drops significantly as the temperature rises (2296.1 MPa @ 30 \degree C to 229.6 MPa @ 120 \degree C) \cite{Zhou2016TemperatureProperties} making the material highly compliant with stretching and distorting.

\begin{figure}[h]
  \centering
  \includegraphics[width=1\linewidth]{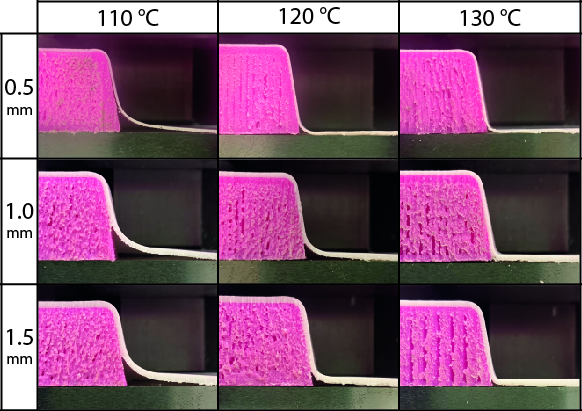}
  \caption{Cross-sectional image of the 3D Printed mold (pink) and vacuum-formed 3D printed sheets (white) for varying thickness and temperature}
  \label{fig:test}
  \Description{
  This figure shows 9 different cross-section photographs of a white 3D printed sheet being vacuum-formed onto a pink 3D printed mold. The images are arranged into 3 by 3 table. From top to below it shows 0.5mm sheet, 1mm sheet and 1.5 mm sheet. From left to right it shows 110 degree Celsius, 120 degree Celsius, and 130 degree Celsius.
  
  }
\end{figure}

Shown in Figure \ref{fig:test}, we vacuum-formed the white 3D printed sheets against the 3D printed mold in pink PLA with a draft angle of 10$\degree$ and 15~mm height. To examine the cross section of the vacuum-formed layer, we used a band saw to cut the samples into halves and measured the void area between the mold and the vacuum-formed layer via pixel counting. We noticed the reduction in the void area between the mold and the 3D printed sheet as the forming temperature increases. Shown in Table \ref{Table_temperature}, we saw a sharp increase in the void area between 0.5~mm and 1.0~mm, but between 1.0~mm and 1.5~mm the change in area was small in comparison. The low-cost desktop vacuum-forming machine was able to heat the 3D printed sheets to the desirable temperature within one minute in the same manner as conventional vacuum-forming.

\begin{table}[h]
 \begin{tabular}{|c| c| c| c|} 
 \hline
Thickness (mm) & 110 \degree C & 120 \degree C & 130 \degree C
\\
 \hline\hline
 0.5 & 5.6 $mm^2$ & 1.6 $mm^2$ & 1.13 $mm^2$ \\ 
 \hline
 1.0 & 25.3 $mm^2$ & 8.9 $mm^2$ & 2.5 $mm^2$
\\
 \hline
 1.5 & 24.1 $mm^2$ & 10.3 $mm^2$ & 4.0 $mm^2$
 \\
 \hline
\end{tabular}
 \caption{Cross-sectional void area after vacuum-forming. The size of mold was 30~mm $\times$ 30~mm $\times$ 15~mm (tall) with draft angle of 10\degree}
\label{Table_temperature}
\Description{
This table goes with the figure 3. it shows the resulting void areas that are formed between the 3D printed sheet and the mold. The table is arranged in 3 by 3 also.

}
\end{table}

\subsection{Colors and 3D patterns}

We first explored the vacuum-formability of the 3D printed sheet with various colored PLA and patterns to construct glowing signage. Vacuum-forming is a common technique for fabricating various glowing signage at the output is lightweight and thin. To prepare 3D printed sheets with multiple colors using single extruder 3D printer, we used a modified tool changing G-code to swap the filament at different points in the print. Unlike the traditional vacuum-forming, 3D printing the sheet allow us to vary the thickness within a single sheet. Therefore, we can have 3D texts, shapes, and braces that are fully infilled and mechanically strong. The thickness variation can also be programmed and 3D printed to counter the thinning from stretch and distort such that the vacuum-formed object has an even thickness distribution after vacuum-forming \cite{Eksi2020ThermoformingSheets}.

\begin{figure}[h]
  \centering
  \includegraphics[width=1\linewidth]{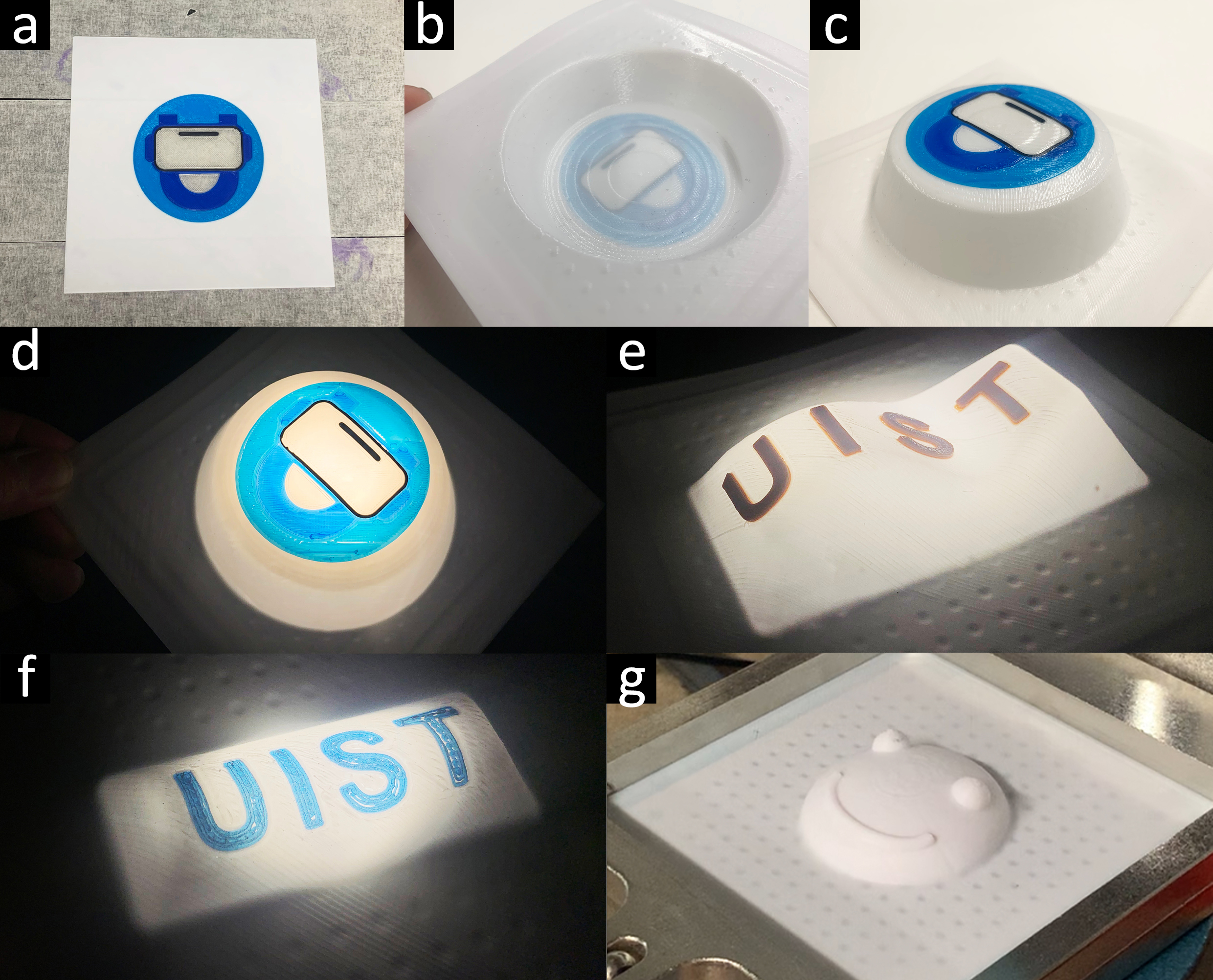}
  \caption{a) 3D printed multi-color sheet as printed; b) Bottom face of the vacuum-formed sheet; c) Top face of the vacuum-formed sheet; d) Glowing 3D UIST signage; e~\&~f) Curvy glowing logo with the 3D text; g) 3D smiley face}
  \label{fig:signage}
  \Description{
  This figure shows a glowing signages constructed by vacuum-forming. (a) shows a 3D printed UIST logo in a flat status. (b) shows a vacuum-formed UIST logo from below. (c) Shows a vacuum-formed logo from the top. The sheet is vacuum formed in a shape of tapered cylinder. (d) shows the glowing signage. (e) shows a different glowing signage vacuum-formed onto curvy surface. (f) shows signage on single curved surface. (g) shows a vacuum-formed smile face.
  }
\end{figure}

\subsection{Vacuum-forming Conductive PLA}

\begin{figure}[h]
  \centering
  \includegraphics[width=1\linewidth]{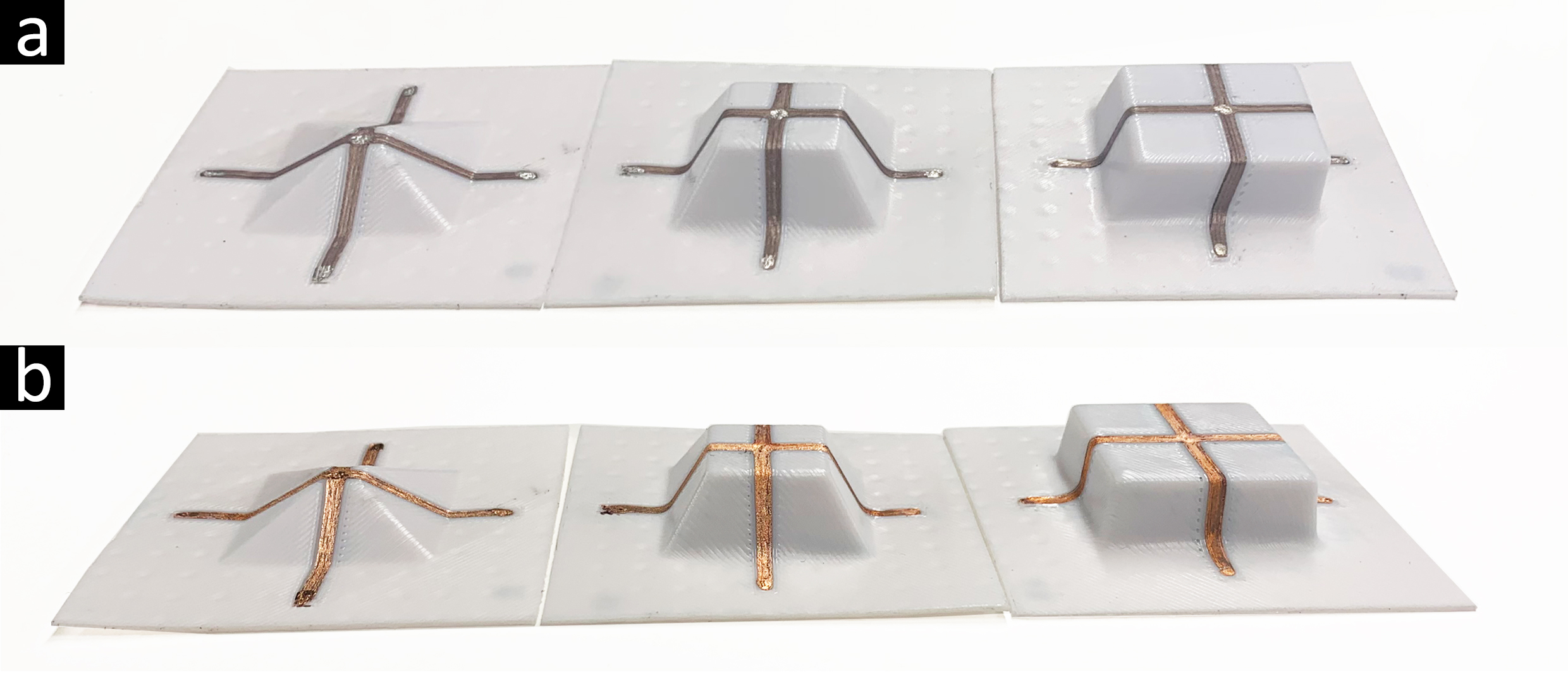}
  \caption{a) Left to right, vacuum-formed 3D printed conductive traces for draft angles of 60\degree, 30\degree, and 4\degree; b) Electroplated vacuum-formed traces}
  \label{fig:vacuumform trace}
  \Description{
  This figure shows a total 6 images of vacuum-formed conductive trace on various angle. (a) shows un-plated conductive trace with a draft angle of 60 degree, 30 degree and 4 degree. (b) shows copper electroplated trace with a draft angle of 60 degree, 30 degree and 4 degree.
  
  }
\end{figure}

We investigated the change in the resistance of the printed conductive PLA before and after vacuum-forming, and after electroplating. We first 3D printed the flat sheet consisting of white and conductive PLA. The conductive trace was printed in a cross-shape; 1.5~mm in width, 0.6~mm in height and 25~mm in length from the centre of the cross to one end. The overall thickness of the sheet material was 0.9~mm printed in 0.3~mm layer height. We first measured the resistance of the trace before vacuum-forming, then we vacuum-formed the 3D printed sheet against three different mold with varying draft angle of 60\degree, 30\degree, and 4\degree. The forming temperature for all 3D printed sheets was set to 120 \degree C. After vacuum forming we measured the change in resistance, then electroplated the traces. Figure \ref{fig:vacuumform trace} shows the image of the 3 vacuum-formed samples prior and post electroplating.

\begin{figure}[h]
  \centering
  \includegraphics[width=1\linewidth]{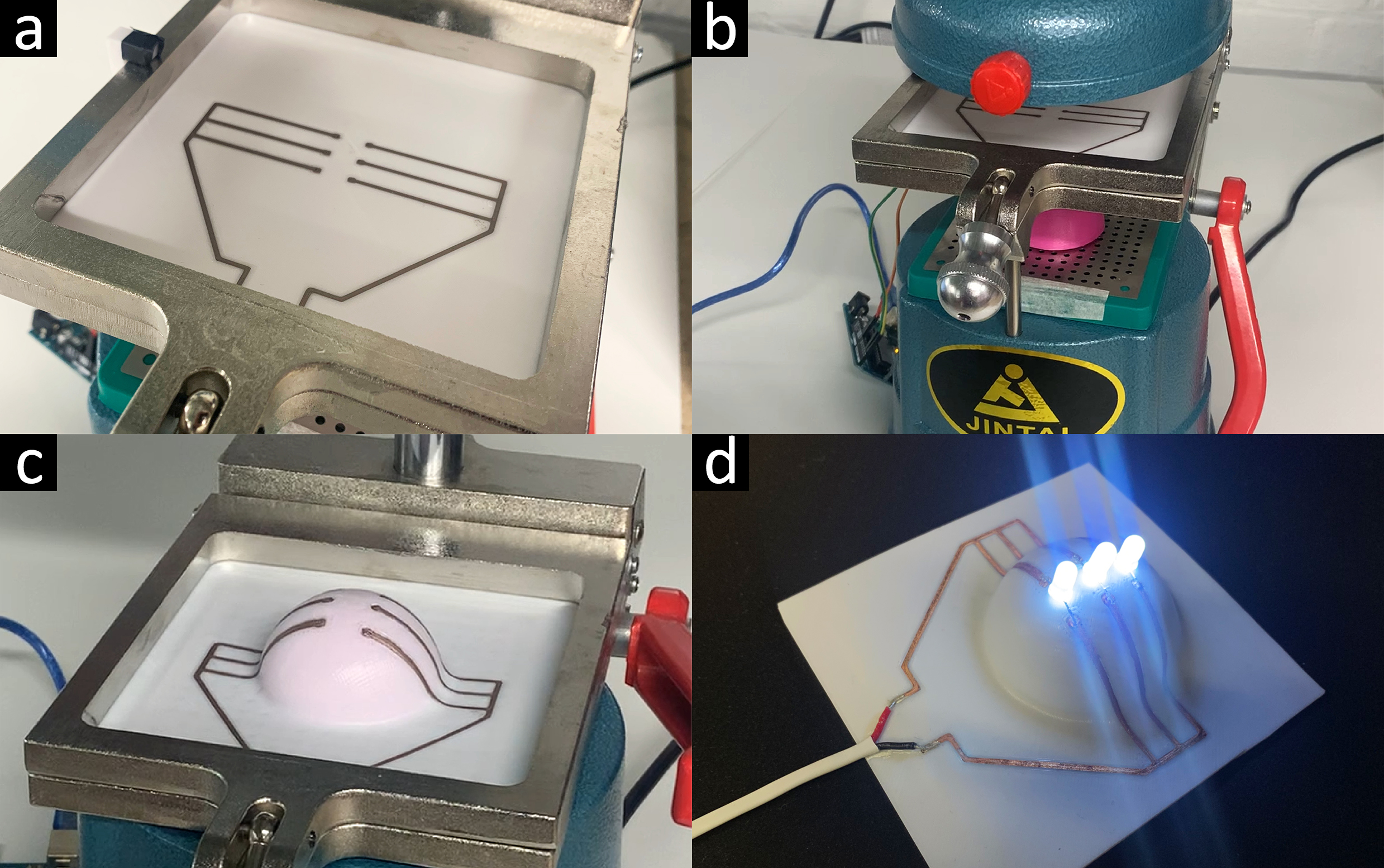}
  \caption{Fabrication process of a dorm display; a) 3D printed vacuum-forming sheet with integrated conductive PLA trace; b) Vacuum-forming set up with the sheet, thermometer and mold; c) Vacuum-formed sheet; d) Electroplated and assembled dorm LED}
  \label{fig:dorm}
  \Description{
  This figure shows a construction process of the dorm-led. (a) shows a vacuum-forming machine with 3D printed sheet clamped onto the machine. The sheet has conductive traces integrated into the white PLA sheet. The trace shows a parallel circuit for lighting 3 LEDs. (b) shows a machine heating the 3D printed sheet on top of 3D printed pink dorm mold. (c) shows a vacuum-formed sheet. It is vacuum-formed onto the hemisphere mold. the conductive trace is also vacuum-formed conformally against the dorm. (d) shows an assembled LEDs circuit with 3 white LED brightly glowing against the dark background.
  }
\end{figure}

As shown in 
Table \ref{Table_resistance}, the resistance of the conductive PLA trace increases as the draft angle decreases, but after electroplating the resistance dropped significantly. Through vacuum-forming the 3D printed sheet, it is possible to construct a doubly-curved, thin and rigid shell-like structure with integrated traces that would be difficult to achieve using conventional planar additive manufacturing. In planar construction the layer structure will be different at every angle, causing inconsistent mechanical properties of the substrate in addition to inconsistent resistance of the printed traces. Figure \ref{fig:dorm} shows an example of a 3D printed and vacuum-formed interconnect on a hemisphere (50~mm diameter) with LEDs mounted.

\begin{table}
 \begin{tabular}{||c| c| c| c||} 
 \hline
Draft Angle
& \multicolumn{1}{p{1.5cm}}{\centering As Printed \\ {\ohm}/cm}
& \multicolumn{1}{|p{1.5cm}}{\centering Vacuum Formed \\ {\ohm}/cm}
& \multicolumn{1}{|p{1.5cm}||}{\centering Electroplated \\ {\ohm}/cm}
\\
 \hline\hline
 60 & 0.0248 & 0.0329 & 0.0015 \\ 
 \hline
 30 & 0.0261 & 0.0465 & *0
\\
 \hline
 4 & 0.024 &0.0575 & 0.0007
 \\
 \hline
\end{tabular}
\caption{Volume resistivity of the conductive PLA trace before vacuum-forming, after vacuum-forming and after electroplating. *The measured resistance of the electroplated trace was below what the multimeter could measure (0.1~\ohm) and thus registered as 0~\ohm}
\label{Table_resistance}
\Description{
Table 2 goes with the figure 5. It shows changes in volume-resistivity of the conductive trace, as printed, vacuum-formed and electroplated for different draft angles.
}
\end{table}


\section{Design Tool}

We used a common 3D modelling software called \textit{Rhino3D} \cite{Rhinoceros3D} with integrated visual scripting extension and live physics engine, namely \textit{Grasshopper} and \textit{Kangaroo} \cite{Grasshopper,KangarooPhysics} respectively, to simulate the vacuum-forming process and design the conformal interconnects. Our simulation allows users to design the interconnects with consideration for the distortion and stretching that occurs during the forming process. The design tool can draw and edit the conformal interconnects conveniently via graphical interface. It then exports the final layout in an STL format ready for 3D printing.

\subsection{Vacuum-forming simulation}
\begin{figure}[h]
  \centering
  \includegraphics[width=1\linewidth]{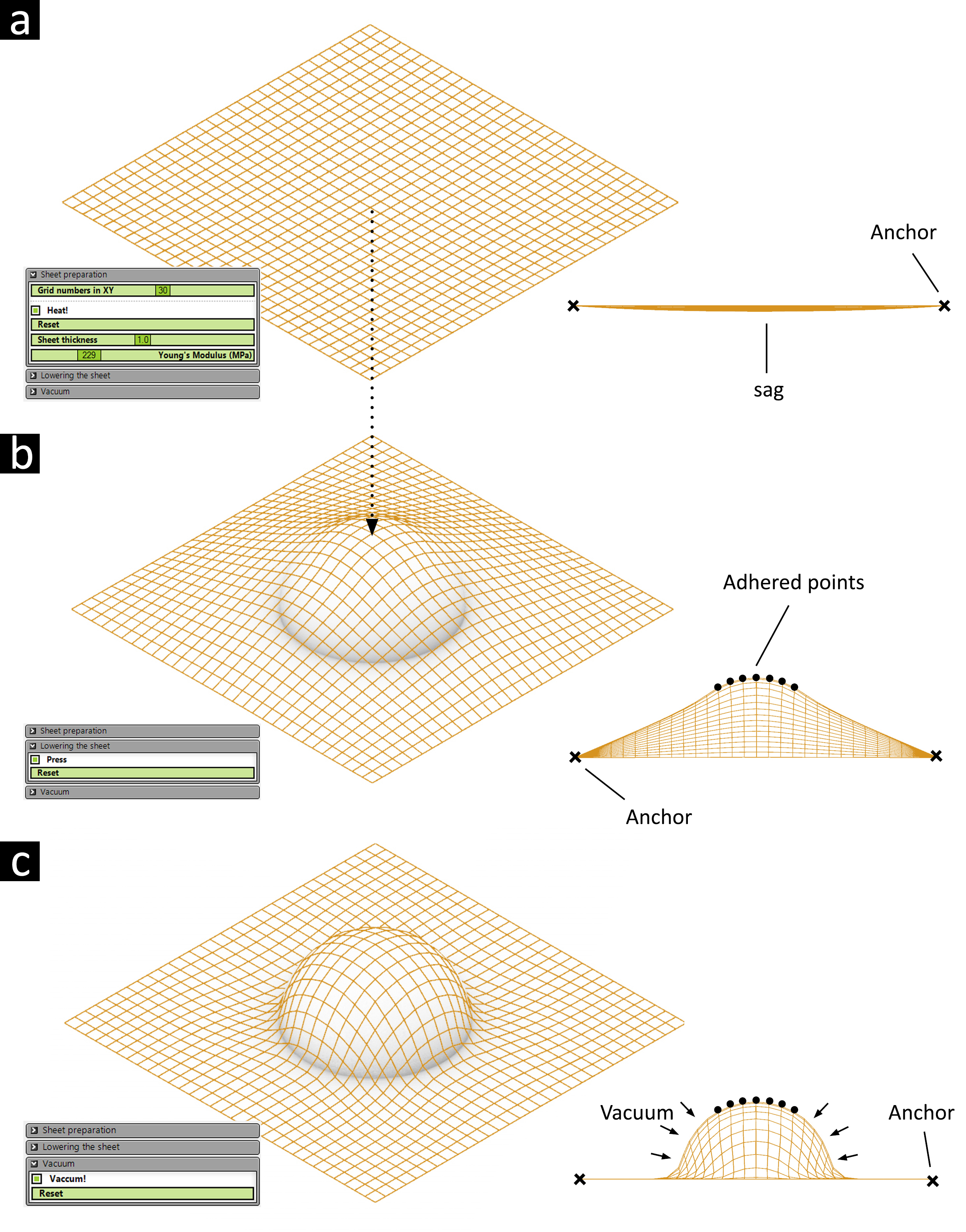}
  \caption{Vacuum-forming stages: a) Preparing the vacuum-forming sheet by heating the sheet for initial sagging; b) Lowering the sheet to the mold whereupon the sheet is stretched against the mold and adhering points are set; c) Vacuum-forming the sheet against the mold, where the vertices are pulled toward the mold.}
  \label{fig:simu}
  \Description{
  This figure shows an illustration of the vacuum-forming process. (a) shows a mesh surface being heated and slightly sagged in the centre. (b) shows a pressed sheet against the dorm mold. the illustration also shows the points that are adhered onto the mold and anchored around the edge. (c) shows a vacuum-formed mesh grid against the mold. 
  }
\end{figure}

Simulation begins with importing a 3D mold into the CAD environment. The user can then define the resolution of the simulation by adjusting the size of the mesh grid and the number of vertices. Complex circuit design will require smaller mesh face and larger quantity of vertices for finer tessellation. Our design tool shows how many vertices there are in the sheet, which also influences the speed of the simulation. 
The simulation consists of three stages: i) heating the sheet; ii) pressing the sheet against the mold, and iii) vacuuming the sheet to the mold. In this simulation we make several assumptions that allow the design software to run the simulation much faster in exchange for minor reduction in accuracy in comparison to the industrial vacuum-forming simulators \cite{Accuform}. Our design software is built in a much simplified manner, specifically aimed for drawing surface interconnects. Our assumptions for vacuum-forming simulations are: i) the thickness of the sheet is not considered, and instead we only consider the overall weight of the sheet, ii) heat is transferred onto the sheet evenly throughout the surface, although in reality the heat transfers more to the central area of the sheet than to the peripheries leading to uneven spread of the tensile strength within the sheet, and iii) adhesion between the mold and the sheet is considered infinite in this simulation, which means once the mesh collides with the mold, the vertices are fixed onto its position and does not slide. The overall process of simulation is illustrated in Figure~\ref{fig:simu}.

\textbf{Heating.} (Figure~\ref{fig:simu}a) The process starts by setting the weight of the sheet, the Young's modulus of PLA at forming temperature, and the resolution of the grid. \textit{Kangaroo} simulates the thermal relaxation of the sheet material by taking the edges of the quadmesh as an axial spring between each vertice. This means that the sheet material is considered as a net of rubber lines-like objects which together approximate the behaviour of a continuous sheet material. \textit{Kangaroo} uses the spring constant ($k =\frac{EA}{L}$) between each vertice, where $E$ is the Young's modulus of the PLA in pascals at forming temperature, $L$ is the length between the vertices, and $A$ is the cross-sectional area. The axial load from gravity is also exerted on every vertice in the negative Z direction. We also select the naked vertices at the edge of the sheet and anchor their position in their current XYZ with an infinite force which assumes that the clamping on vacuum-forming machine does not let the sheet slide at all during the entire process. In our design editor the user can press the button for heating, which begins the simulation.


\textbf{Pressing.} (Figure~\ref{fig:simu}b) In this process the sheet is lowered to the mold and the bed. During the iteration process Kangaroo checks if the sheet's mesh vertices are in collision with the mold. If the vertices are in collision with the mold, those vertices are kept outside the input mold and it's position remains fixed. In this simulation the mold is regarded as input only, which means that the mold is not impacted by the collision. In this process we anchor the XY position of the vertices at the edge of the mesh.

\textbf{Vacuum.} (Figure~\ref{fig:simu}c). For the vacuum-forming process, we first anchor the vertices that are already in contact with the mold and also the vertices at the edge of the mesh that are clamped. Then, we pull the vertices toward the mold. In this process no other external forces are considered.

\subsection{Calibration}

\begin{figure}[h]
  \centering
  \includegraphics[width=1\linewidth]{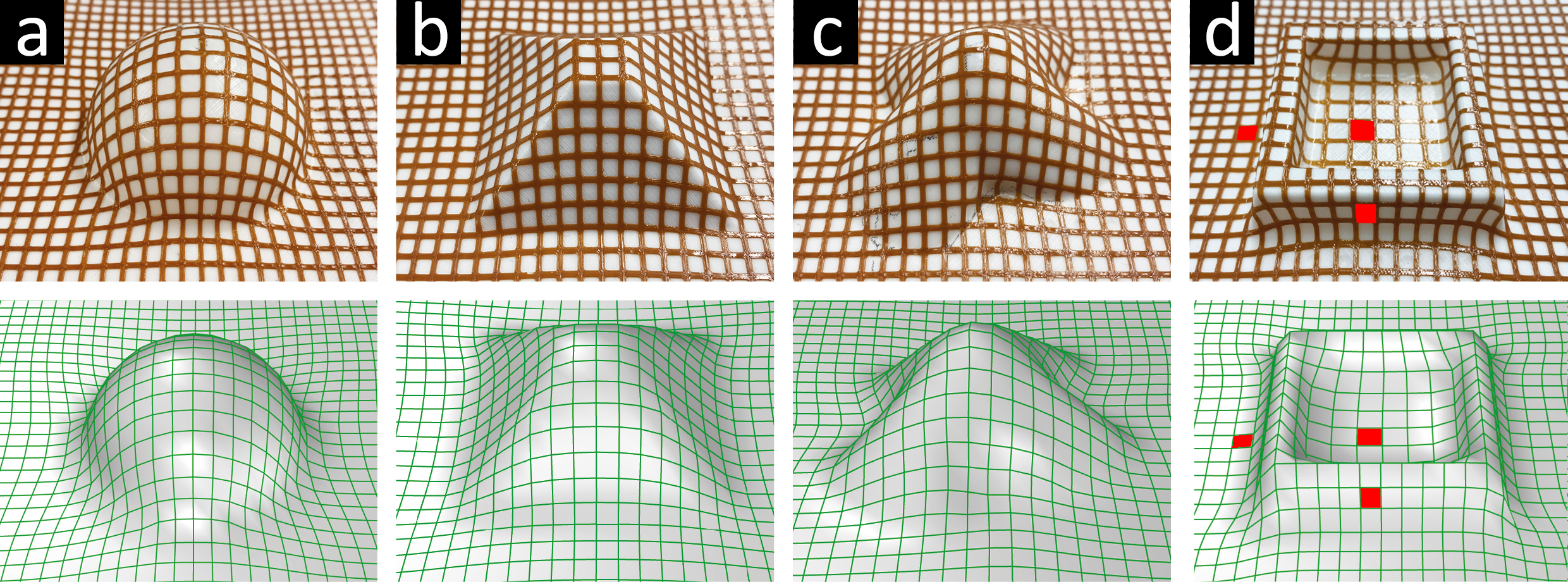}
  \caption{Vacuum-forming shapes for calibration: a) sphere, b) pyramid, c) curvy, and d) concave.}
  \label{fig:cali}
  \Description{
  This figure compares what is actually vacuum-formed against the simulated model. All the photographs (a-d) shows a 3D printed white PLA sheet with brown PLA grid. it also shows a simulated model with green mesh grid. both appear very close. (a) shows a hemisphere, (b) shows pyramid, (c) shows a oddly curved surface, (d) shows a rectangular concave shape with red boxes colored which indicate the reference point for calibration.
  }
\end{figure}

\begin{figure}[h]
  \centering
  \includegraphics[width=0.5\linewidth]{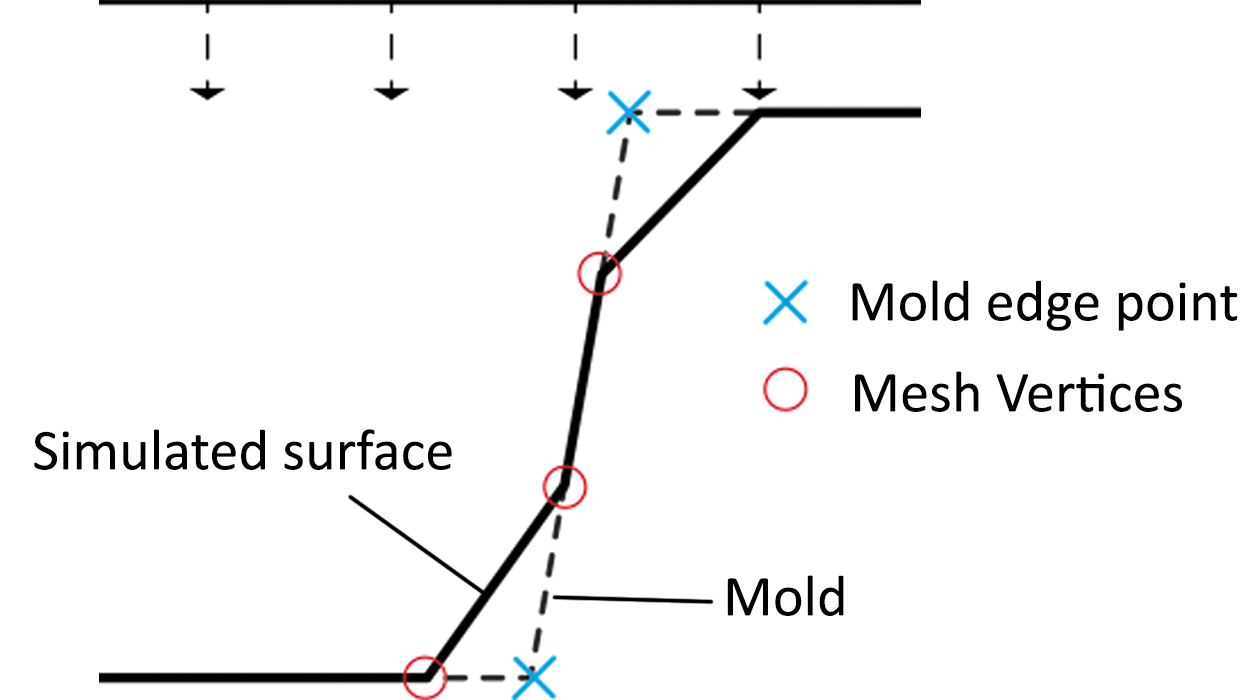}
  \caption{Tessellation of the vacuum-formed mesh vertices: reduced resolution of the simulation can cause the tessellation that may not align with the mold in perfect form}
  \label{fig:cali2}
  \Description{
  Shows a line illustration of the computed model. it shows a vacuum-formed lines being mis-aligned with the mold due to the control points between the mesh being to wide.
  }
\end{figure}

To validate that the simulated digital 3D model matches the actual physical object, we compared the two models in various shapes including hemisphere, frustum, arbitrary curved surface and rectangular concave. We 3D printed the sheet with a brown grid that is equally spaced as the digital mesh in the simulation. We noticed that our simulated model stretched more around the perimeter of the sheet in comparison to the actual vacuum-formed sheet. We find that this is due to our assumption that the heat is evenly spread during the heating process. To measure the accuracy of the simulation we measured the differences in the length of line segments at three locations on the surface of our concave model (red boxes in Figure~\ref{fig:cali}d). The differences were 0.6~mm at the square outside the mold, 0.4~mm at the square on the surface of the draft, and 0.2~mm at inner concave surface. The amount of stretch may vary depending on PLA filament manufacturer, but a user can simply tweak the value of the Young's modulus to adjust the simulation. One issue regarding the simulation is that the mesh vertices may not align perfectly with the edge condition of the mold, and the constructed polygon may appear misaligned (Figure~\ref{fig:cali2}). This can be resolved by increasing the amount of mesh faces but at an expense of reduced simulation speed.

\subsection{Designing interconnects}
\begin{figure}[h]
  \centering
  \includegraphics[width=1\linewidth]{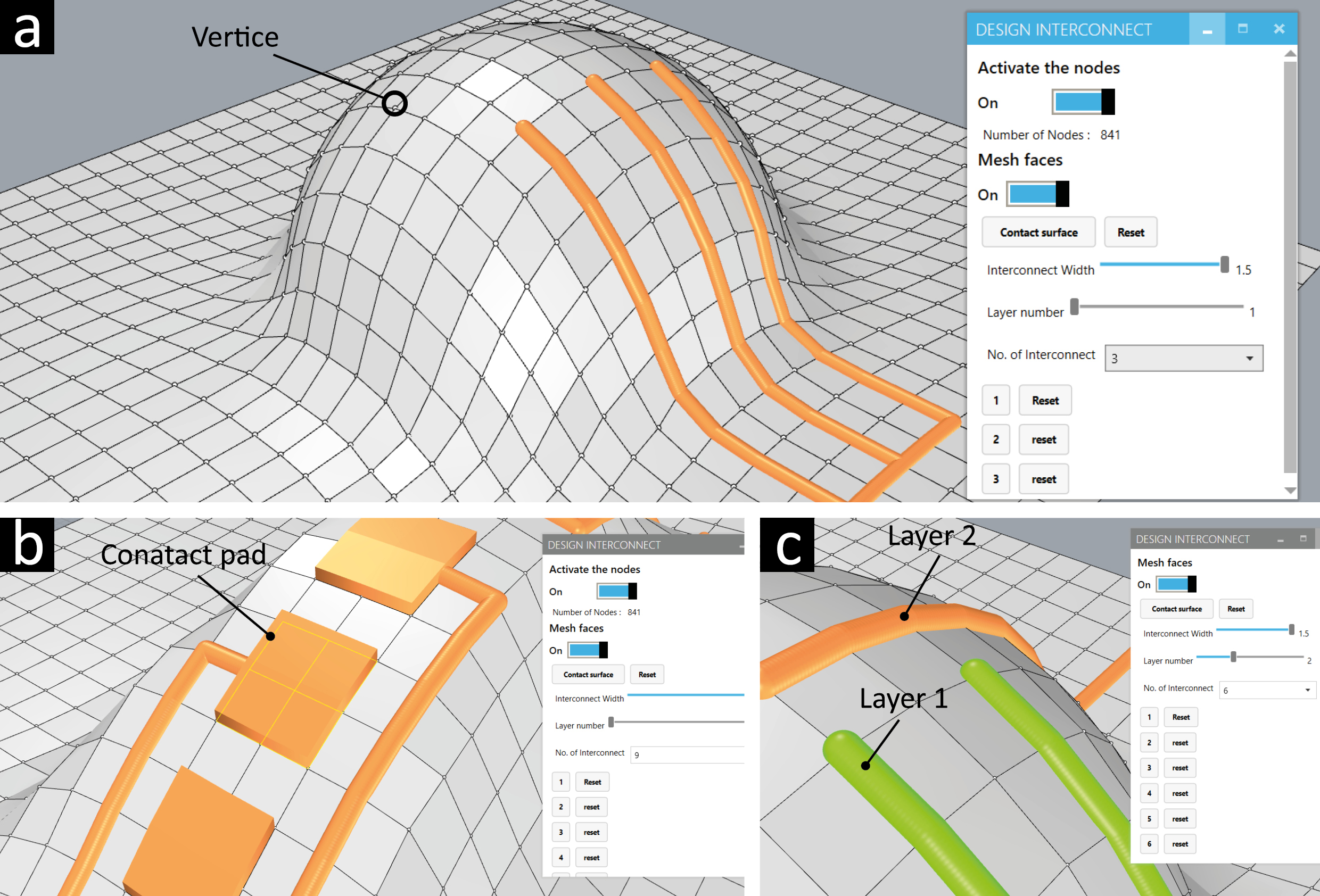}
  \caption{Design interface. a) Designing interconnect by selecting the vertices; b) Designing the contact pad by selecting mesh faces; c) Toggling between different layers to draw the interconnect for different layer number (when switched the interconnects are color coded for visual reference).}
  \label{fig:inter}
  \Description{
  This figure shows a CAD environment for drawing 3D interconnects onto the simulated semi-sphere surface. It shows a vertices that are constructed on the mesh faces. The design editor window allow users to click the vertices to draw the interconnect. The editor window consists of toggles and button. The simulated figure (a) shows a orange interconnect for parallel LED circuit being drawn. (b) Orange buttons are drawn onto the surface. (c) shows green and orange interconnects drawn. the colours represents the layer numbers.
  }
\end{figure}

Once the simulation is complete, a user can start designing the circuit on the surface using our design editor. A user can simply select the vertices on the surface of the vacuum-formed sheet to draw interconnects (Figure~\ref{fig:inter}a). A user can also select the mesh faces to create electrical contacts (Figure~\ref{fig:inter}b). A user can adjust the width of the interconnects by changing the value on slider at any point during the design process. The design editor uses color coding to differentiate the interconnects for different layers (Figure~\ref{fig:inter}c). 

\begin{figure}[h]
  \centering
  \includegraphics[width=1.0\linewidth]{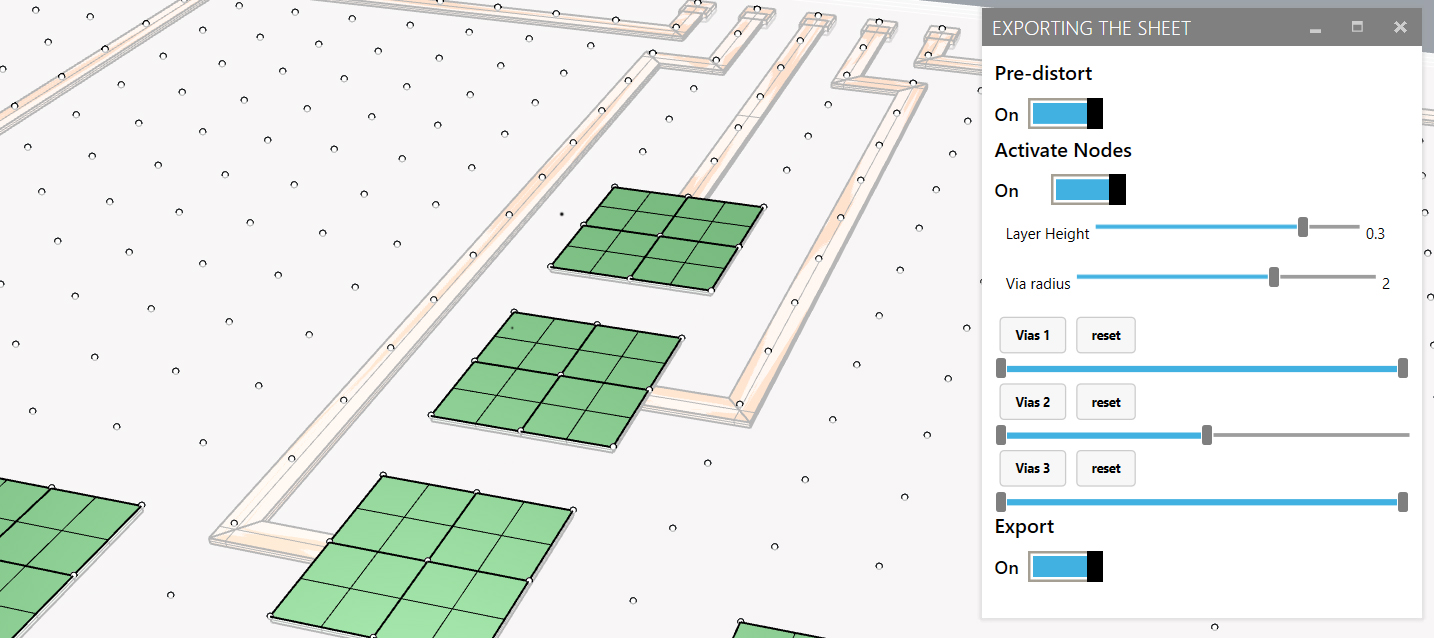}
  \caption{Final stage of the design editor. (Left) Projected conductive pattern in planar surface. (Right) Design editor allowing users to vary layer height and the size and location of vias prior to exporting}
  \label{fig:flat}
  \Description{
  This figure shows a CAD environment with flat sheet model with buttons and interconnects. button are colored in green and interconnects are colored in orange. The design editor window shows a toggle, buttons and slider which control the width and height of the interconnects and electrical contact. This window also allow user to export the model. 
  }
\end{figure}

Once the design is complete, a user can pre-distort the design into planar sheet (Figure \ref{fig:flat}. In this process, the user can change the layer height, create vias that connect the interconnects between the layers. Once completed, a user can then export the geometry in STL, which can be sliced for 3D printing.

\section{Example Applications}

To illustrate the unique characteristics of the vacuum-formed circuit boards, we fabricated three examples that incorporate sensing and actuation. As discussed earlier, 3D printing the sheet materials and vacuum-forming allows designers to customise the circuit by selectively exposing the contact areas and embedding the traces between PLA layers.

\subsection{Smart Tray}
\begin{figure}[h]
  \centering
  \includegraphics[width=1\linewidth]{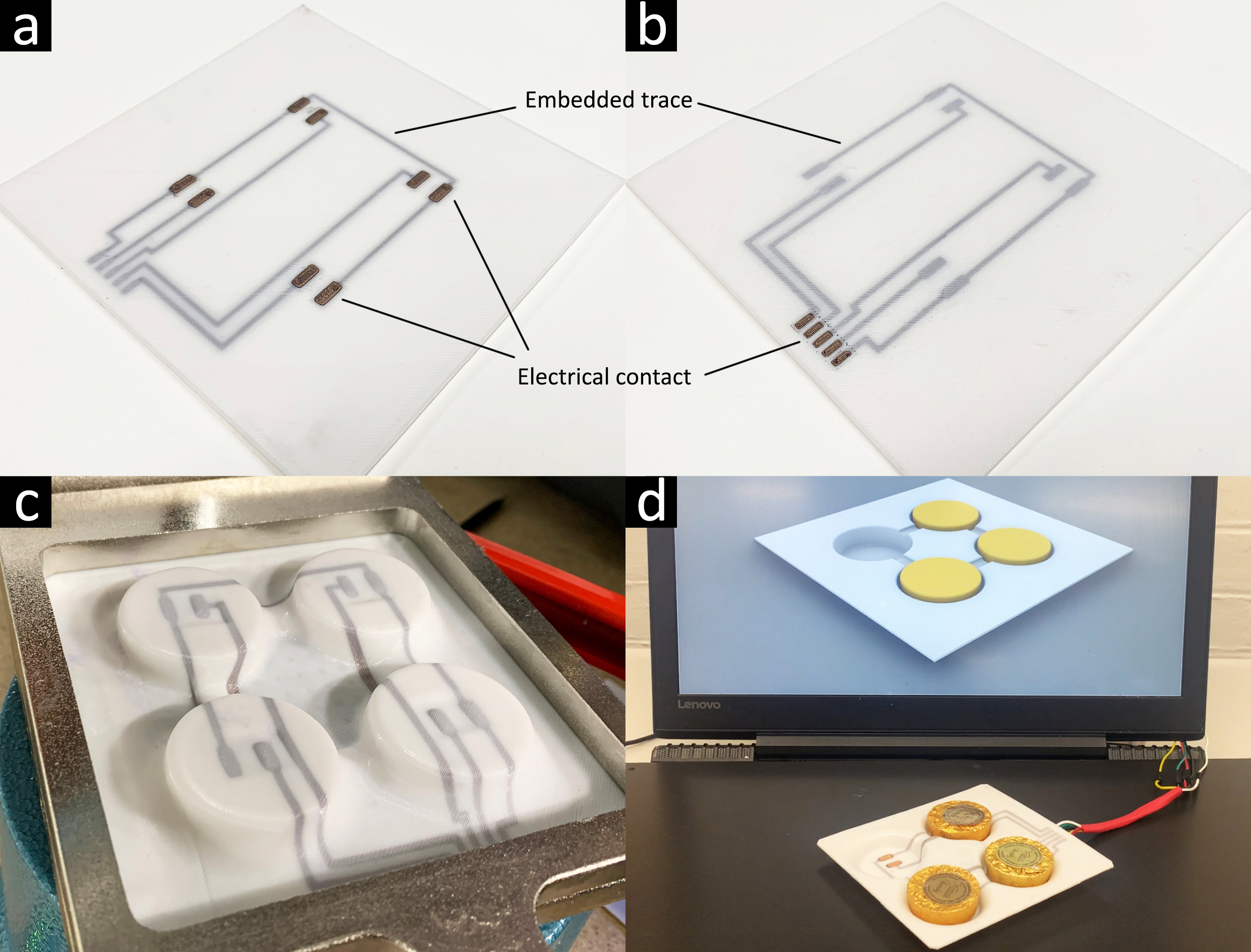}
  \caption{Fabrication process of the smart tray. a) Top face of the 3D printed sheet; b) Bottom face of the 3D printed sheet, thermometer and mold; c) Vacuum-formed sheet against the mold; d) Assembled tray connected to digital screen visually animating the tray and the chocolates located within.}
  \label{fig:tray}
  \Description{
  Shows construction process of the smart tray. (a) and (b) shows 'as printed' sheet from top and bottom. This sheet has embedded interconnects and exposed electrical contacts. (c) shows a vacuum-formed 3D printed sheet on the machine. (d) shows a working smart tray connected to computer which identifies how many chocolate is being placed on tray. 
  }
\end{figure}

Vacuum-formed trays and packages are extensively used in consumer goods owing to light weight and rigidness that makes them ideal for deliveries and storage. In this example, we show a smart tray with surface interactivity (Figure~\ref{fig:tray}). The smart tray has exposed conductive pads inside the concave surface of the tray which detect the presence of an object. At bottom face of the sheet there are electrical contacts for wiring. Beside the contacts that are needed to be exposed, all the printed traces are insulated between the PLA layers. 

\subsection{3D Buttons and Sliders}
\begin{figure}[t]
  \centering
  \includegraphics[width=1\linewidth]{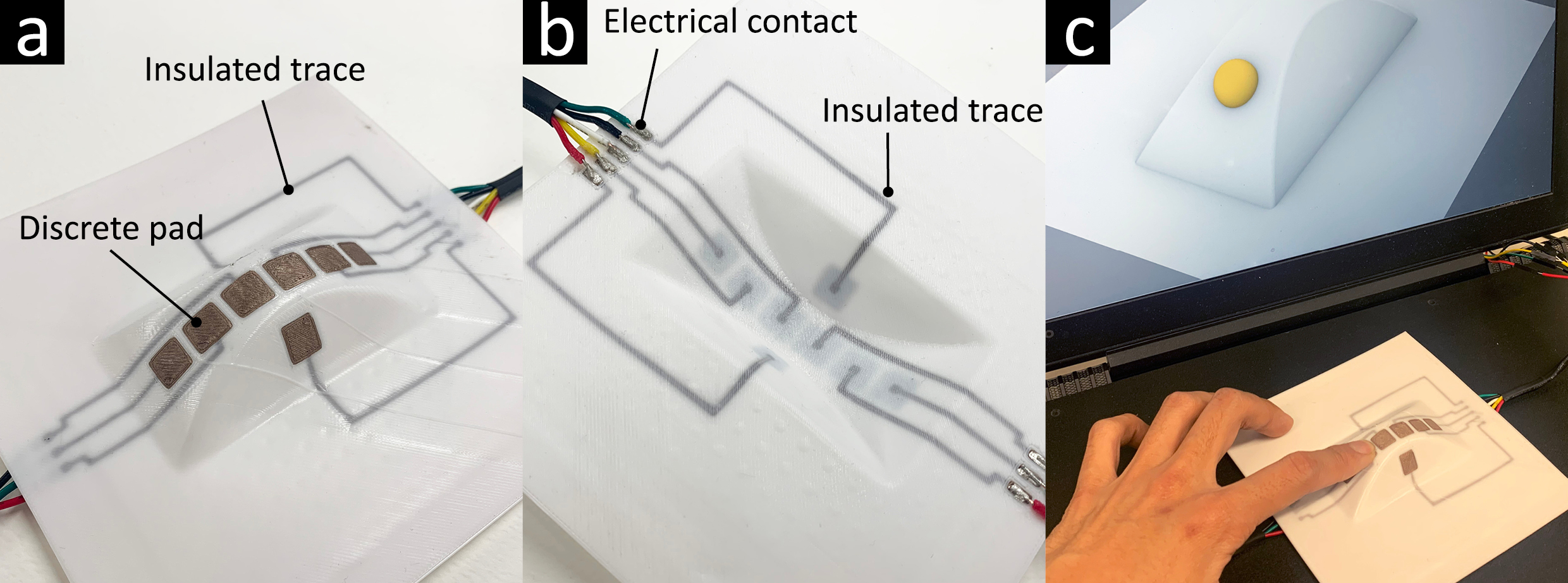}
  \caption{a) Top face of the conformal slider; b) Bottom face of the conformal slider; c) Conformal slider connected to digital screen with visual animation}
  \label{fig:slider}
  \Description{
  3D printed and vacuum-formed buttons and sliders. (a) shows a top surface of the device which consists of discrete buttons on curved surface with wires connected. (b) shows a bottom face of the vacuum-formed device with embedded conductive traces and exposed electrical contacts. (c) shows working device connected to computer which shows a sphere that shows the location of the finger tip on the device.
  
  }
\end{figure}

Embedding conductive interconnects within a thin and freeform surface is desirable for smart interactive surfaces. Embedded circuits allow miniaturization of electrical arrangements by eliminating unnecessary spaces for attaching separate PCBs and wires. Our vacuum-formed circuit board approach can construct embedded traces for doubly curved surfaces, which increase freedom of design. In this example we show a doubly curved, shell-like sensing device consisting of discrete sliders and buttons  (Figure~\ref{fig:slider}).

\subsection{Modular Lamp}
\begin{figure}[h]
  \centering
  \includegraphics[width=1\linewidth]{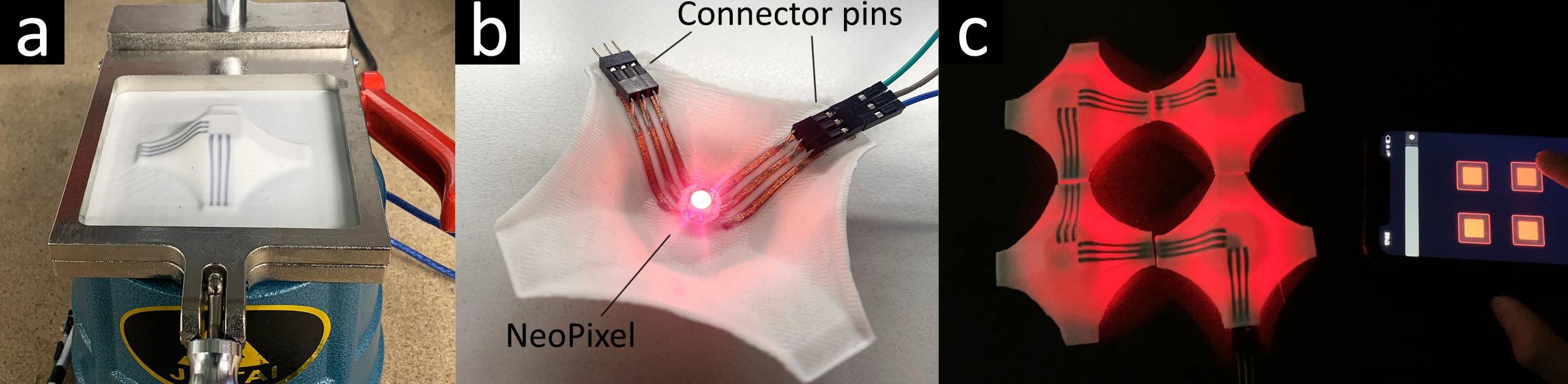}
  \caption{a) Vacuum-formed 3D printed sheet; b) Electroplated and assembled modular light with NeoPixel and connector pins; c) Assembled units controlled via smartphone}
  \label{fig:mod}
  \Description{
  This figure shows a construction process of modular light. (a) vacuum-formed 3D printed sheet on the vacuum-forming machine. The product is shaped in Celtic-cross. (b) shows the bottom face the sheet with interconnect and the button LED. The image shows three connecting pins. (c) Red modular light connected to iPhone. iphone controls the individual light unit. 
  
  }
\end{figure}

One of the key benefits of vacuum-forming is that since the same mold can be re-used multiple times, it can repetitively and accurately reproduce the same objects in a convenient way. Here we present a modular lamp (Figure~\ref{fig:mod}) controlled via a smartphone. For the modular lamp, we placed conductive traces to the bottom face of the sheet and electroplated it for minimum resistance.

\section{Discussion}

In this section, we highlight some limitations of the vacuum-formed circuit boards and potential future directions.

\textbf{Sensitive print settings}: In our experiments we noticed that the quality of vacuum-formed 3D printed sheet is sensitive to the printer settings. For instance, if the PLA is under-extruded, the strands within the sheet can de-laminate and create holes during the vacuum-forming process. This is because small air gaps in 3D printed sheet widen during the forming process. The 3D printed sheet, therefore, needs to be printed with a higher extrusion rate to ensure that there is sufficient overlapping between each printing path to avoid any air gaps.

\textbf{Limited multi-material}: While attempting to mix various materials together in a single sheet, we noticed that due to the differences in the tensile properties of each materials, de-lamination can occur at junctions where two or more materials overlap. This was particularly the case when trying to mix conventional PLA with thermoplastic polyurethane (TPU) filament. Since TPU has a much stronger elasticity, the stretching and distortion differed significantly, causing detachments during vacuum-forming.

\textbf{PLA constraints}: Currently, the variety of commercially available conductive filaments for FDM 3D printers is limited. The currently available conductive filaments have too high resistance. While \textit{Electrifi} from Multi3D has significantly lower resistance in comparison to other competitors, it is very sensitive to heat. During the heating process of the sheet, the resistance of the 3D printed trace increases significantly due to thermal annealing \cite{Cardenas2020FlashThermoplastics}. Although the exposed traces can be copper electroplated, the embedded traces cannot be plated and are thus most useful for sensing signals. The method of vacuum-forming 3D printed circuit and 3D printed electronics in general could benefit from having more diverse conductive filaments with improved thermal performance and conductivity.

\section{Conclusion}

In this paper we presented a method of vacuum-forming 3D printed sheets to construct thin, rigid and freeform interactive surfaces with embedded interconnects. We demonstrated the manufacturing technique using only  low-cost and highly accessible equipment that individual designers can employ in a home environment. We first explored the `vacuum-form-ability' of 3D printed sheets with embedded conductive traces and evaluated their electrical performance for different draft angles. We showcased range of examples that are unique to our manufacturing technique. We also demonstrated a common 3D CAD environment with integrated physics engine to simulate the vacuum-forming for various mold shapes and provided a design editor for drawing interconnects onto conformal surfaces. Our design editor conveniently pre-distorts the design into the planar sheet ready for 3D printing. Hybrid manufacturing techniques are continuing to expand capabilities and opportunities for personal fabrication and manufacturing at the edge. Our work aims to contribute to the field of personal fabrication by exploring new manufacturing techniques for 3D printed electronics with unique from factors and characteristics.

\balance{}
\bibliographystyle{ACM-Reference-Format}
\bibliography{references}

\end{document}